\documentclass[12pt]{article}
\usepackage{amsmath,epsfig,graphicx,verbatim} 
\include{epsf} 
\def\ltap{\raisebox{-.6ex}{\rlap{$\,\sim\,$}} \raisebox{.4ex}{$\,<\,$}} 
\def\gtap{\raisebox{-.6ex}{\rlap{$\,\sim\,$}} \raisebox{.4ex}{$\,>\,$}}

\newcommand\as{\alpha_{\mathrm{S}}} 
\newcommand\f[2]{\frac{#1}{#2}} 
\def\beq{\begin{equation}} 
\def\eeq{\end{equation}} 
\def\beeq{\begin{eqnarray}} 
\def\eeeq{\end{eqnarray}} 
 
\def\to{\rightarrow}
\def\ito{\leftarrow} 
\def\nn{\nonumber} 
 
\def\tL{{\tilde L}}
 
\def\qt{q_T}

\begin{document} 
\begin{titlepage}
\renewcommand{\thefootnote}{\fnsymbol{footnote}}
\vspace*{2cm}

\begin{center}
{\Large \bf Production of Drell--Yan lepton pairs
in hadron collisions:\\[0.2cm]
transverse-momentum resummation \\[0.4cm] 
at
next-to-next-to-leading logarithmic accuracy}
\end{center}

\par \vspace{2mm}
\begin{center}
{\bf Giuseppe Bozzi${}^{(a)}$,
Stefano Catani${}^{(b)}$,}\\ 
\vskip .2cm
{\bf Giancarlo Ferrera${}^{(b)}$,
Daniel de Florian${}^{(c)}$
and
Massimiliano Grazzini${}^{(b)}$}\\
\vspace{5mm}
${}^{(a)}$Dipartimento di Fisica, Universit\`a di Milano and\\ INFN, Sezione di Milano,
I-20133 Milan, Italy\\
%
${}^{(b)}$INFN, Sezione di Firenze and
Dipartimento di Fisica, Universit\`a di Firenze,\\
I-50019 Sesto Fiorentino, Florence, Italy\\
${}^{(c)}$Departamento de F\'\i sica, FCEYN, Universidad de Buenos Aires,\\
(1428) Pabell\'on 1 Ciudad Universitaria, Capital Federal, Argentina\\
\vspace{25mm}
\end{center}

\par \vspace{2mm}
\begin{center} {\large \bf Abstract} \end{center}
\begin{quote}
\pretolerance 10000

We consider the transverse-momentum ($q_T$)
distribution 
of Drell--Yan lepton pairs produced
in
hadron collisions. 
At small values of $q_T$, we resum the logarithmically-enhanced 
perturbative QCD contributions
up to next-to-next-to-leading logarithmic accuracy. 
At intermediate and large values
of $q_T$,
we consistently combine resummation with the 
known next-to-leading order
perturbative result.
All perturbative terms up to order $\as^2$ are included in our computation 
which, after integration over $q_T$, reproduces
the known next-to-next-to-leading order result for the Drell--Yan 
total cross section. 
We show and discuss the reduction 
in the scale dependence of the results with respect to
lower-order calculations, estimating
the corresponding perturbative uncertainty.
We present a preliminary comparison with
Tevatron Run~II data.

\end{quote}

\vspace*{\fill}
\begin{flushleft}
July 2010

\end{flushleft}
\end{titlepage}

\setcounter{footnote}{1}
\renewcommand{\thefootnote}{\fnsymbol{footnote}}
\section{Introduction}


The hadroproduction of the vector bosons $W$ and $Z/\gamma^*$,
also known as the Drell--Yan (DY) process~\cite{Drell:1970wh},
is the process to which parton model ideas
(previously developed for deep inelastic lepton--hadron scattering)
were first applied in the context of hard-scattering processes
in hadron--hadron collisions.

At high-energy hadron colliders, such as the Tevatron and the LHC,
vector bosons are produced with large rates and with 
relatively-simple experimental signatures.
The vector boson production process is thus relevant for various reasons.
It is important for detector calibration; it provides us with strong tests
of perturbative QCD and, in particular, it gives stringent information
on the 
parton densities
of the colliding hadrons;
it represents an important background for new-physics searches.
Owing to these reasons,
it is essential to have accurate theoretical predictions for
vector boson production cross sections and related kinematical distributions.

These predictions are based on perturbative QCD and are obtained
as power series expansions 
in the 
strong coupling $\as$.
The total cross section \cite{Hamberg:1990np}
and the rapidity distribution of the vector boson \cite{Anastasiou:2003ds}
are known up to the next-to-next-to-leading order (NNLO)
in QCD perturbation theory.
Fully exclusive NNLO calculations, including the leptonic decay 
of the vector boson, 
are also available \cite{Melnikov:2006di,Catani:2009sm,Catani:2010en}.
Electroweak corrections up to ${\cal O}(\alpha)$
have been computed
both for $W$ \cite{ewW} and $Z/\gamma^*$ production~\cite{ewZ}.

In this paper we consider the transverse-momentum ($q_T$) spectrum
of the vector boson.
The $q_T$ spectra of the $W$ and $Z$ bosons are particularly important since
the uncertainties in their shape directly affect
the measurement of the $W$ mass.
In the large-$q_T$ region ($q_T\sim m_V$), where the transverse momentum is
of the order of the vector boson mass $m_V$,
perturbative QCD
calculations based on the truncation of the perturbative series at a
fixed order in $\as$ 
are theoretically justified.
In this region, the QCD radiative corrections are known up to the next-to-leading
order (NLO) \cite{Ellis:1981hk,Arnold:1988dp,Gonsalves:1989ar}.
Nonetheless 
the bulk of the vector boson events is produced in the small-$q_T$ region ($q_T\ll m_V$),
where the convergence of the fixed-order
expansion is spoiled by the presence of 
large logarithmic terms, $\as^n\ln^m (m^2_V/q_T^2)$.
To obtain reliable predictions, 
these logarithmically-enhanced terms 
have to be systematically
resummed to all
perturbative orders \cite{Dokshitzer:hw}--\cite{Catani:2000vq}.
%
The resummed and fixed-order
calculations 
at small and large values of 
$q_T$ can then be consistently matched at intermediate values of $q_T$,
to obtain QCD predictions for the entire range of transverse momenta.

We use the transverse-momentum resummation formalism proposed 
in Refs.~\cite{Catani:2000vq,Bozzi:2005wk}. 
The formalism is valid for 
a generic process in which
a high-mass system of non strongly-interacting particles is produced 
in hadron--hadron collisions.
The method has so far been applied to the production of
the Standard Model (SM) Higgs boson 
\cite{Bozzi:2003jy, Bozzi:2005wk, Bozzi:2007pn},
single vector bosons
\cite{Bozzi:2008bb},
$WW$ \cite{Grazzini:2005vw} and $ZZ$ \cite{Frederix:2008vb} pairs,
slepton pairs \cite{Bozzi:2006fw}, and
DY lepton pairs in polarized collisions \cite{Jiro}.
The study of Ref.~\cite{Bozzi:2008bb} is mainly based on 
next-to-leading logarithmic (NLL) resummation at small
$q_T$ and on the leading-order (LO) calculation at large $q_T$.
In this paper we extend the analysis and the results of Ref.~\cite{Bozzi:2008bb}, combining
the most advanced  perturbative information that is available at present:
next-to-next-to-leading logarithmic (NNLL) resummation  at small
$q_T$ and the NLO calculation at large $q_T$.
Other phenomenological studies of the vector boson
$q_T$ distribution, which combine resummed and fixed-order perturbative
results at 
various levels of theoretical accuracy,
can be found in Refs.~\cite{Arnold:1990yk}. 
%
%

The paper is organized as follows. In Sect.~\ref{sec:theory} we briefly review 
the resummation formalism of Refs. \cite{Catani:2000vq,Bozzi:2005wk} 
and its application
to vector boson production.
In Sect.~\ref{sec:results} 
we present numerical results
for $Z/\gamma^*$ production, 
and we 
comment on their comparison 
with the Tevatron Run~II data \cite{:2007nt,Abazov:2010kn}.
We also study the scale dependence of our results to the purpose of
estimating the corresponding
perturbative uncertainty. 
In Sect.~\ref{sec:summa} we summarize our results.

\section{Transverse-momentum resummation}
\label{sec:theory}

We briefly recall some of the 
main points of the transverse-momentum resummation formalism of
Refs.~\cite{Catani:2000vq,Bozzi:2005wk}.
Here we consider the specific case of DY
lepton pair production,
i.e. the production
of a vector boson $V$ ($V=W^+,W^-,Z/\gamma^*$) that subsequently decays in a
lepton pair. 

The inclusive hard-scattering process is
\begin{equation}
h_1(p_1) + h_2(p_2) \;\to\; V (M,q_T ) + X \;\to\; l_1+l_2 + X,   
\label{first}
\end{equation}
where $h_1$ and $h_2$ are the colliding hadrons with momenta
$p_1$ and $p_2$, $V$ is the vector boson
(which 
decays in the
lepton pair $l_1,l_2$)
with invariant mass
$M$ and transverse momentum $q_T$,
and $X$ is an arbitrary and undetected final state. 

According to the QCD factorization theorem
the $q_T$ 
differential cross 
section $d\sigma_V/dq_T^2$ can be written as
\begin{equation}
\label{dcross}
\f{d\sigma_V}{d q_T^2}(q_T,M,s)= \sum_{a,b}
\int_0^1 dx_1 \,\int_0^1 dx_2 \,f_{a/h_1}(x_1,\mu_F^2)
\,f_{b/h_2}(x_2,\mu_F^2) \;
\f{d{\hat \sigma}_{V ab}}{d q_T^2}(q_T, M,{\hat s};
\as(\mu_R^2),\mu_R^2,\mu_F^2) 
\;\;,
\end{equation}
where $f_{a/h}(x,\mu_F^2)$ ($a=q,{\bar q}, g$)
are the parton densities of the colliding hadron $h$ 
at the factorization scale $\mu_F$, 
$d\hat\sigma^V_{ab}/d{q_T^2}$ are the perturbative QCD 
partonic cross sections, 
$s$ ($\hat s = x_1 x_2 s$) 
is the square of the 
hadronic (partonic) centre--of--mass  energy, 
and $\mu_R$ is the renormalization scale. 

In the region where 
$q_T \sim  M$ (in practice, we always consider the case in which $M$ is close
to the mass $m_V$ of the vector boson),
the QCD perturbative
series is controlled by a small expansion parameter, 
$\as(M)$,
and fixed-order calculations are
theoretically justified. In this region, 
the QCD radiative corrections are known up to next-to-leading order 
(NLO)~\cite{Ellis:1981hk}. 

In the small-$q_T$ region 
($q_T\ll M$),
the convergence of the fixed-order
perturbative expansion is spoiled
by the presence 
of powers of large logarithmic terms, 
$\as^n\ln^m (M^2/q_T^2)$.
To obtain reliable predictions these terms have to be resummed to all orders.

We perform the resummation
at the level of the partonic cross section, which
is decomposed~as
\begin{equation}
\label{resplusfin}
\f{d{\hat \sigma}_{V\,ab}}{dq_T^2}=
\f{d{\hat \sigma}_{V\,ab}^{(\rm res.)}}{dq_T^2}
+\f{d{\hat \sigma}_{V\,ab}^{(\rm fin.)}}{dq_T^2}\; .
\end{equation}
The first term on the right-hand side
contains all the logarithmically-enhanced contributions,
which have to be resummed
to all orders in $\as$,
while the second 
term 
is free of such contributions and
can thus be evaluated at fixed order in perturbation theory. 
Using the Bessel transformation between the conjugate variables 
$q_T$ and $b$ ($b$ is the impact parameter),
the resummed component $d{\hat \sigma}^{({\rm res.})}_{V\,ab}$
can be expressed as
\begin{equation}
\label{resum}
\f{d{\hat \sigma}_{V \,ab}^{(\rm res.)}}{dq_T^2}(q_T,M,{\hat s};
\as(\mu_R^2),\mu_R^2,\mu_F^2) 
=\f{M^2}{\hat s} \;
\int_0^\infty db \; \f{b}{2} \;J_0(b q_T) 
\;{\cal W}_{ab}^{V}(b,M,{\hat s};\as(\mu_R^2),\mu_R^2,\mu_F^2) \;,
\end{equation}
where $J_0(x)$ is the $0$th-order Bessel function.
Considering
the Mellin $N$-moments ${\cal W}_N$ of ${\cal W}$ with respect to the variable 
$z=M^2/{\hat s}$ at fixed $M$,
the resummation structure of ${\cal W}_{ab, \,N}^V$ can 
be 
organized in exponential 
form~\footnote{For the sake of simplicity we consider here only
the case of 
the diagonal terms in the flavour space 
of the partonic indices $a,b$. For the general case and a 
detailed discussion of the resummation formalism, we refer to  
Ref.~\cite{Bozzi:2005wk}.}
\begin{align}
\label{wtilde}
{\cal W}_{N}^{V}(b,M;\as(\mu_R^2),\mu_R^2,\mu_F^2)
&={\cal H}_{N}^{V}\left(M, 
\as(\mu_R^2);M^2/\mu^2_R,M^2/\mu^2_F,M^2/Q^2
\right) \nonumber \\
&\times \exp\{{\cal G}_{N}(\as(\mu^2_R),L;M^2/\mu^2_R,M^2/Q^2
)\}
\;\;,
\end{align}
were we have defined the logarithmic expansion parameter $L\equiv \ln ({Q^2 b^2}/{b_0^2})$,
and $b_0=2e^{-\gamma_E}$ ($\gamma_E=0.5772...$ 
is the Euler number).
The scale $Q$ ($Q\sim M$), 
which appears on the right-hand side of Eq.~(\ref{wtilde}),
is the resummation scale \cite{Bozzi:2005wk}. Although ${\cal W}_{N}^{V}$ (i.e., the product
${\cal H}_{N}^{V} \times \exp\{{\cal G}_{N}\}$) does not depend on $Q$ when
evaluated to all perturbative orders, its explicit dependence on $Q$
appears when ${\cal W}_{N}^{V}$ is computed by truncation of the resummed
expression at some level of logarithmic accuracy (see Eq.~(\ref{exponent})
below). 
Variations of $Q$ around $M$ can thus be used to estimate the
size of yet uncalculated 
higher-order
logarithmic contributions.

The universal\footnote{The form factor does not depend on the type
of produced vector boson. More generally, all the hard-scattering processes 
initiated
by quark-antiquark annihilation have the same form factor.} 
form factor $\exp\{ {\cal G}_N\}$
contains all
the terms that order-by-order in $\as$ are logarithmically divergent 
as $b \to \infty$ (or, equivalently, $q_T\to 0$). 
The resummed logarithmic expansion of the exponent ${\cal G}_N$ 
is defined as follows:
\begin{align}
\label{exponent}
{\cal G}_{N}(\as, L;M^2/\mu^2_R,M^2/Q^2)&=L \;g^{(1)}(\as L)+g_N^{(2)}(\as L;M^2/\mu_R^2,M^2/Q^2)\nn\\
&+\f{\as}{\pi} g_N^{(3)}(\as L,M^2/\mu_R^2,M^2/Q^2)+\dots
\end{align}
where the term $L\, g^{(1)}$ collects the leading logarithmic (LL) 
contributions, the function $g_N^{(2)}$ includes
the next-to-leading leading logarithmic
(NLL) contributions \cite{Kodaira:1981nh}, 
$g_N^{(3)}$ controls the NNLL 
terms
\cite{Davies:1984hs, Davies:1984sp, deFlorian:2000pr}
and so forth. The explicit form of the functions
$g^{(1)}$, $g_N^{(2)}$ and $g_N^{(3)}$ can be found in Ref.~\cite{Bozzi:2005wk}.
The process dependent function ${\cal H}_N^{V}$ 
does not depend on the impact parameter $b$ and it 
includes all the perturbative
terms that behave as constants as 
$b \to \infty$. 
It can thus be expanded in powers of $\as=\as(\mu_R^2)$:
\begin{align}
\label{hexpan}
{\cal H}_N^{V}(M,\as;M^2/\mu^2_R,M^2/\mu^2_F,M^2/Q^2)&=
\sigma_{V}^{(0)}(M)
\Bigl[ 1+ \f{\as}{\pi} \,{\cal H}_N^{V \,(1)}(M^2/\mu^2_F,M^2/Q^2) 
\Bigr. \nn \\
&+ \Bigl.
\left(\f{\as}{\pi}\right)^2 
\,{\cal H}_N^{V \,(2)}(M^2/\mu^2_R,M^2/\mu^2_F,M^2/Q^2)+\dots \Bigr] \;\;,
\end{align}
where $\sigma_{V}^{(0)}$ is the partonic cross section at the Born level.
The first-order coefficients ${\cal H}_{q{\bar q}\ito ab,N}^{V(1)}$
in Eq.~(\ref{hexpan})
are known since a long time \cite{Davies:1984hs},
while
the second-order coefficients ${\cal H}_{q{\bar q}\ito ab,N}^{V(2)}$
were computed only recently~\cite{Catani:2009sm}.

Within a straightforward (`naive') implementation of Eq.~(\ref{wtilde}), 
the resummation of the large logarithmic contributions 
would affect
not only the small-$q_T$ region, but also the region of large values of $q_T$.
This can easily be understood by observing that the logarithmic expansion 
parameter $L$ 
diverges also
when $b\to 0$.
To reduce the impact of unjustified higher-order contributions in the large-$q_T$ region,
the logarithmic variable $L$ in Eq.~(\ref{wtilde})
is actually replaced 
by $\tL\equiv \ln \left({Q^2 b^2}/{b_0^2}+1\right)$ 
\cite{Bozzi:2005wk, Bozzi:2003jy}.
This replacement has an additional and relevant
consequence: after inclusion of the finite component (see Eq.~(\ref{resfin})),  
we exactly recover the fixed-order perturbative value of the total cross section
upon integration of the $q_T$  distribution over $q_T$
(i.e., the resummed terms give a vanishing contribution upon integration over $q_T$).

We now turn to consider
the finite component of the transverse-momentum cross section
(see Eq.~(\ref{resplusfin})).
Since $d\sigma_V^{({\rm fin.})}$ does not contain large logarithmic terms
in the small-$q_T$ region,
it can be evaluated by truncation of the perturbative series
at a given fixed order.
In practice, the finite component is computed starting from the usual
fixed-order perturbative truncation of the partonic cross section and
subtracting the expansion of the resummed part at the same perturbative order.
Introducing the subscript f.o. to denote the perturbative truncation of the
various terms, we have:
\begin{equation}
\label{resfin}
\Bigl[ \f{d{\hat \sigma}_{V \,ab}^{(\rm fin.)}}{d q_T^2} \Bigr]_{\rm f.o.} =
\Bigl[\f{d{\hat \sigma}_{V \,ab}^{}}{d q_T^2}\Bigr]_{\rm f.o.}
- \Bigl[ \f{d{\hat \sigma}_{V \,ab}^{(\rm res.)}}{d q_T^2}\Bigr]_{\rm f.o.} \;.
\end{equation} 
This matching procedure 
between resummed and finite contributions guarantees
to achieve uniform theoretical accuracy 
over 
the region from small to intermediate values of transverse momenta. 
At large values of $q_T$,
the resummation (and matching) procedure is eventually superseded by the
customary fixed-order calculations 
(their theoretical accuracy in the large-$q_T$ region cannot be
improved by resummation of the logarithmic terms that dominate 
in the small-$q_T$ region).

In summary,
the inclusion of the functions $g^{(1)}$, $g_N^{(2)}$,
${\cal H}_N^{V(1)}$ in the resummed component,
together with the evaluation of the finite component at LO (i.e. at ${\cal O}(\as)$),
allows us to perform the resummation at NLL+LO accuracy.
This is the theoretical accuracy used in our previous study \cite{Bozzi:2008bb}
of the DY $q_T$ distribution.
Including also the functions $g_N^{(3)}$ and ${\cal H}_N^{V(2)}$, together 
with the finite component at NLO (i.e. at ${\cal O}(\as^2)$)
leads to full NNLL+NLO accuracy.
The perturbative coefficient $A^{(3)}$, which contributes to the NNLL
function $g_N^{(3)}$ (see, e.g., Eq.~(24) in Ref.~\cite{Bozzi:2005wk}), 
is not yet known. 
In the following,
we assume that the value of $A^{(3)}$
is the same as the one \cite{Vogt:2000ci, Moch:2004pa} that appears in 
resummed calculations of soft-gluon 
contributions near partonic threshold.
Using the recently computed ${\cal H}_N^{V(2)}$ 
coefficient~\cite{Catani:2009sm}, 
we are thus able to present the complete result for the
DY $q_T$-distribution up to NNLL+NLO accuracy.
We 
point out
that the NNLL+NLO (NLL+LO) result includes the {\em full} NNLO (NLO)
perturbative contribution in the small-$\qt$ region.
In particular,  the NNLO (NLO) result for the total cross section  
is exactly recovered upon integration
over $q_T$ of the differential cross section $d \sigma_V/dq_T$ at NNLL+NLO
(NLL+LO) accuracy.

We conclude this section with some comments on the numerical implementation
of our calculation.
Within our formalism, the resummation factor ${\cal W}_{N}^{V}(b,M)$
is directly defined, at fixed $M$,
in the space of the conjugate variables $b$ and $N$. 
To obtain the hadronic cross
section,
we have to perform inverse integral
transformations: the Bessel transformation in Eq.~(\ref{resum}) and an inverse
Mellin transformation.
These integrals
are carried out numerically. The Mellin inversion 
requires the numerical evaluation
of some basic $N$-moment functions that appear in the expression of the  
the second-order coefficients ${\cal H}_{q{\bar q}\ito ab,N}^{V(2)}$
\cite{Catani:2009sm}: this evaluation 
has to be performed 
for complex values of N, and we use
the numerical results of Ref.~\cite{Blumlein:2000hw}.
We recall \cite{Bozzi:2005wk} that the resummed form factor 
$\exp \{{\cal G}_N(\as(\mu_R^2),{\widetilde L})\}$
is singular at 
the values of $b$ where $\as(\mu_R^2) {\widetilde L} \geq \pi/\beta_0$ 
($\beta_0$ is the first-order coefficient of the QCD $\beta$ function). 
Performing the 
Bessel
transformation with respect to the impact parameter $b$ 
(see Eq.~(\ref{resum})), we deal with this
singularity as 
we did
in Ref.~\cite{Bozzi:2005wk},
by using the regularization prescription of
Refs.~\cite{Laenen:2000de,Kulesza:2002rh}:
the singularity is avoided by deforming the 
integration contour in the complex $b$ space.

\section{Numerical results for $Z/\gamma^*$ production at the Tevatron}
\label{sec:results}

In this section 
we consider $Z/\gamma^*$ production in $p{\bar p}$ collisions at Tevatron
energies. We present our resummed results at NNLL+NLO accuracy,
we compare them with the NLL+LO results 
(the NLL+LO results
in Ref.~\cite{Bozzi:2008bb} were obtained by using the
MRST2004 NLO parton densities \cite{Martin:2004ir}),
and we comment on the comparison 
with Tevatron Run~II data \cite{:2007nt,Abazov:2010kn}. 

The hadronic $q_T$ cross section at NNLL+NLO (NLL+LO) accuracy
is computed by using the MSTW2008 NNLO (NLO) parton 
densities \cite{Martin:2009iq}, 
with $\as(\mu_R^2)$ evaluated at 3-loop (2-loop) order.
This choice of the order of the parton densities and $\as$
is fully justified both in the small-$q_T$ region
(where the calculation of the partonic cross section includes the complete
NNLO (NLO) result and is controlled by NNLL (NLL) 
resummation) and in the intermediate-$q_T$ region
(where the calculation is  
constrained by the value of the NNLO (NLO) total cross section).

As for the electroweak couplings, we use the so called $G_\mu$ scheme,
where the input parameters are $G_F$ , $m_Z$, $m_W$.
In particular, we use the PDG 2008 \cite{Amsler:2008zzb} values
$G_F = 1.16637\times 10^{-5}$~GeV$^{-2}$,
$m_Z = 91.1876$~GeV, $\Gamma_Z=2.4952$~GeV, $m_W = 80.398$~GeV.
Our calculation implements 
the decays $\gamma^* \to l^+l^-$ and $Z^* \to l^+l^-$
at fixed value of the invariant mass of the $l^+l^-$ pair. In particular, we
include the effects of the $\gamma^*\,Z$ interference and of the 
finite width of the $Z$ boson.
Nonetheless, 
the numerical results 
presented below
are obtained by simply using the 
narrow-width approximation 
and neglecting the photon contribution. We find that this approximation works 
to better than 1\% accuracy in the inclusive region of lepton invariant
mass that is covered by the D0 data\footnote{The measured $q_T$ spectra
are inclusive over the following regions of lepton invariant mass:
70--110~GeV \cite{:2007nt} and 65--115~GeV \cite{Abazov:2010kn}.}.

As discussed in Sect.~\ref{sec:theory}, the 
resummed calculation depends on the factorization and 
renormalization scales and on the resummation scale $Q$. 
Our convention to compute factorization 
and renormalization scale uncertainties is to consider
independent variations of $\mu_F$ and $\mu_R$ by a factor of two around 
the central values $\mu_F=\mu_R=m_Z$
(i.e. we consider the range $m_Z/2\leq \{\mu_F,\mu_R\}\leq 2\,m_Z$), with the constraint
$0.5 \leq \mu_F/\mu_R \leq 2$. 
Similarly, we follow Ref.~\cite{Bozzi:2008bb} and
we choose $Q=m_Z/2$ as central value of the resummation scale,
considering scale variations in the 
range $m_Z/4 < Q < m_Z$.

\begin{figure}[htb]
\begin{center}
\begin{tabular}{cc}
\includegraphics[width=0.47\textwidth]{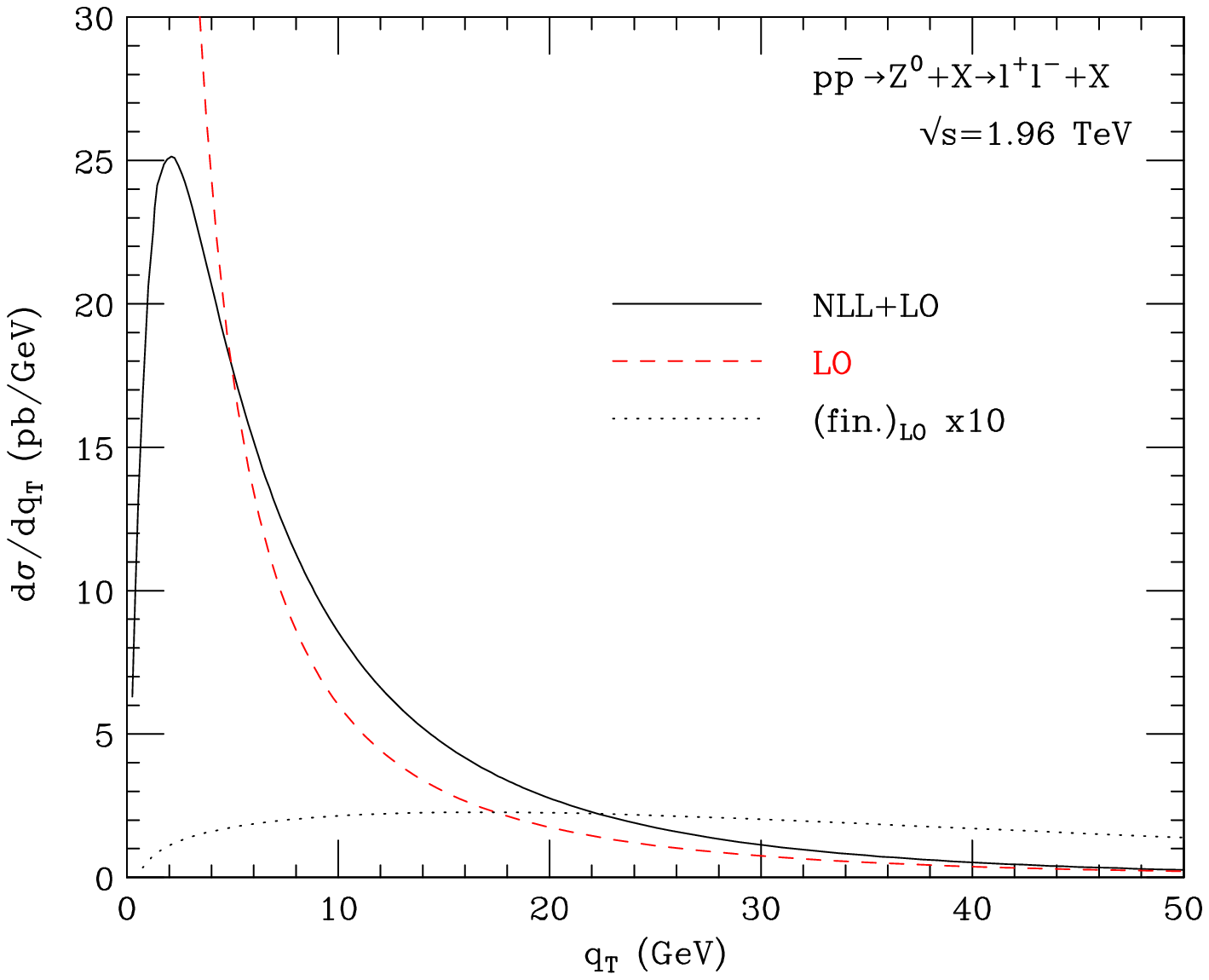} & \includegraphics[width=0.47\textwidth]{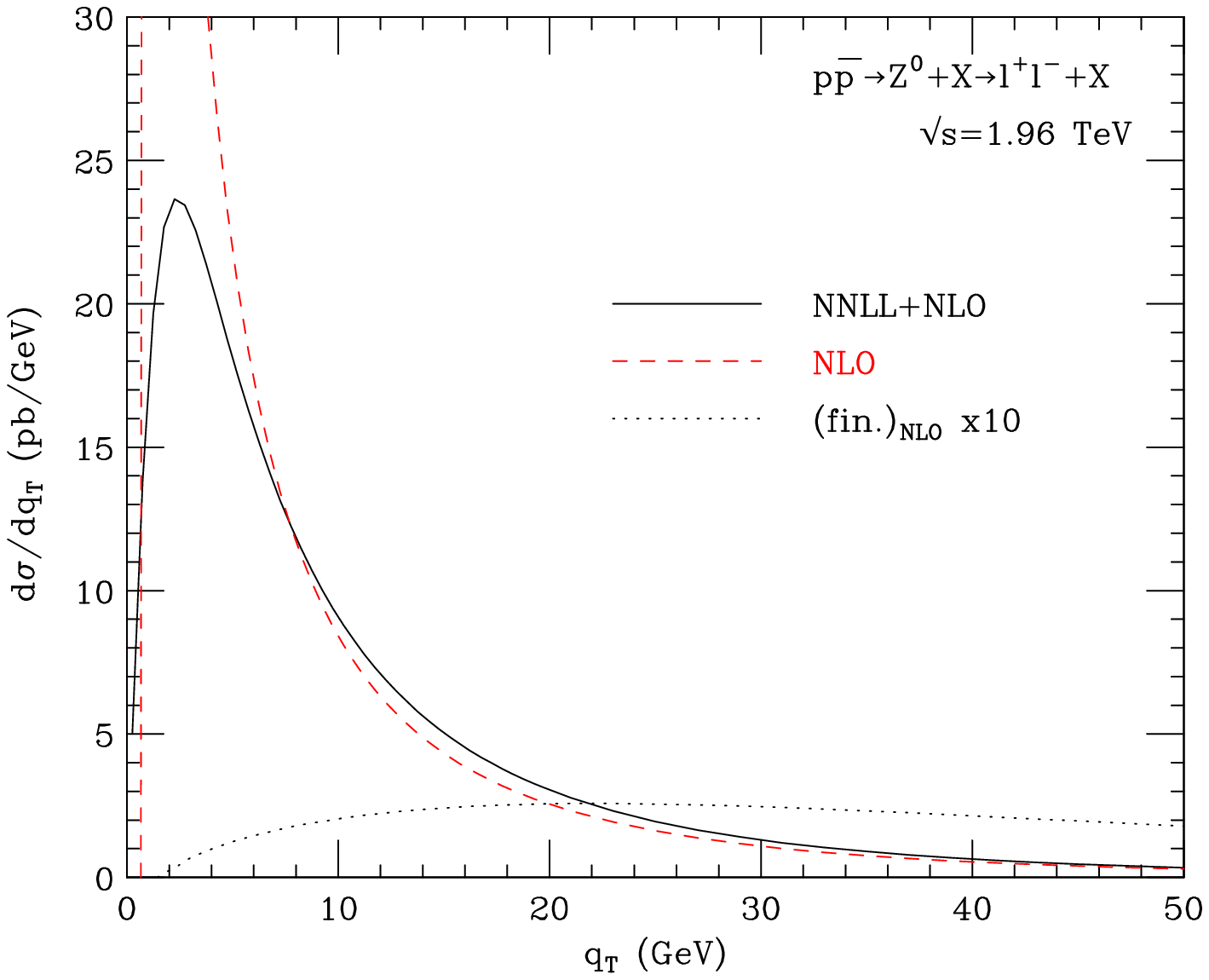}\\
\end{tabular}
\end{center}
\caption{\label{fig1}
{\em The $q_T$ spectrum of $Z$ bosons at the Tevatron Run II: results at NLL+LO (left panel) and NNLL+NLO (right panel)
accuracy. Each result is compared to the corresponding fixed-order result
(dashed line) and to the finite component (dotted line) in Eq.~(\ref{resfin}).}}
\end{figure}


In Fig.~\ref{fig1} (left panel) we present 
the NLL+LO $q_T$ spectrum
at the Tevatron Run~II ($\sqrt{s}=1.96$~TeV)\footnote{Analogous results
at the Tevatron Run~I ($\sqrt{s}=1.8$~TeV),
obtained by using the MRST2004 parton densities \cite{Martin:2004ir}, were presented
in the left panel of Fig.~6 of Ref.~\cite{Bozzi:2008bb}.}.
The NLL+LO 
result (solid line) at
the default scales ($\mu_F=\mu_R=m_Z$, $Q=m_Z/2$) is compared with the
corresponding
LO result (dashed line).
The LO finite component of the spectrum (see Eq.~(\ref{resplusfin})),
multiplied
by a factor of 10 to make it more visible, is also shown 
for comparison (dotted line).
We see that the LO result diverges to $+\infty$ as $q_T\to 0$.
The resummation of the small-$q_T$ logarithms
leads to a well-behaved distribution: it vanishes as $q_T\to 0$, has
a kinematical peak at $q_T\sim 2$~GeV, and tends to the corresponding LO result
at large
values of $q_T$.
The finite component
smoothly vanishes
as $q_T\to 0$ and gives a small contribution to the NLL+LO result in the low-$q_T$ region.

The results in the right panel of Fig.~\ref{fig1}
are analogous to those in the left panel, although systematically at one order
higher. The $q_T$ spectrum at NNLL+NLO accuracy (solid line) is compared with
the NLO result (dashed line) and with the NLO finite component of the spectrum
(dotted line).
The NLO result diverges to $-\infty$ as $q_T\to 0$ and, at small values of $q_T$,
it has an unphysical peak (the top of the peak is above the vertical scale of the plot)
that is produced
by the numerical compensation of negative leading
and positive subleading logarithmic contributions.
The contribution of the NLO finite component to the NNLL+NLO result
is smaller than 1\% at the peak
and becomes more important as $q_T$ increases: it is about 8\% at $q_T\sim
20$~GeV, about 20\% at $q_T\sim 30$~GeV and about 53\% at $q_T\sim 50$~GeV.
A similar quantitative behaviour is observed by considering the contribution of
the NLO finite component to the NLO result;
the contribution is about 10\% at $q_T\sim 20$~GeV, 
about 22\% at $q_T\sim 30$~GeV and about 60\% at $q_T\sim 50$~GeV.
In the region of intermediate values of $q_T$ (say, around 20~GeV),
the difference between the NNLL+NLO and NLO results is larger than the size
of the NLO finite component. This difference is produced by the logarithmic
terms (at NNLO and beyond NNLO) that are included in the resummed calculation at
NNLL accuracy. At large values of $q_T$ the contribution of
the NLO finite component sizeably increases. This behaviour indicates that  
the logarithmic terms are no longer dominant and that the resummed
calculation cannot improve upon the predictivity of the fixed-order expansion. 

Comparing the left and right panels of Fig.~\ref{fig1}, we see that
the $q_T$ spectrum is slightly harder at NNLL+NLO accuracy than at NLL+LO accuracy.
The height of the peak  at NNLL+NLO is lower than at NLL+LO.
This is mainly due to the fact that the NNLO total cross section,
which fixes the value of the $q_T$ integral of our NNLL+NLO result,
is only about 3\% larger than the NLO total cross section, whereas
in the region of intermediate values of $q_T$ the cross section 
at NLO (and, correspondingly, at NNLL+NLO)
is definitely larger than at LO (and, correspondingly, at NLL+LO);
this leads to a reduction of the cross section 
at small $q_T$.


\begin{figure}[htb]
\begin{center}
\begin{tabular}{cc}
\includegraphics[width=0.48\textwidth]{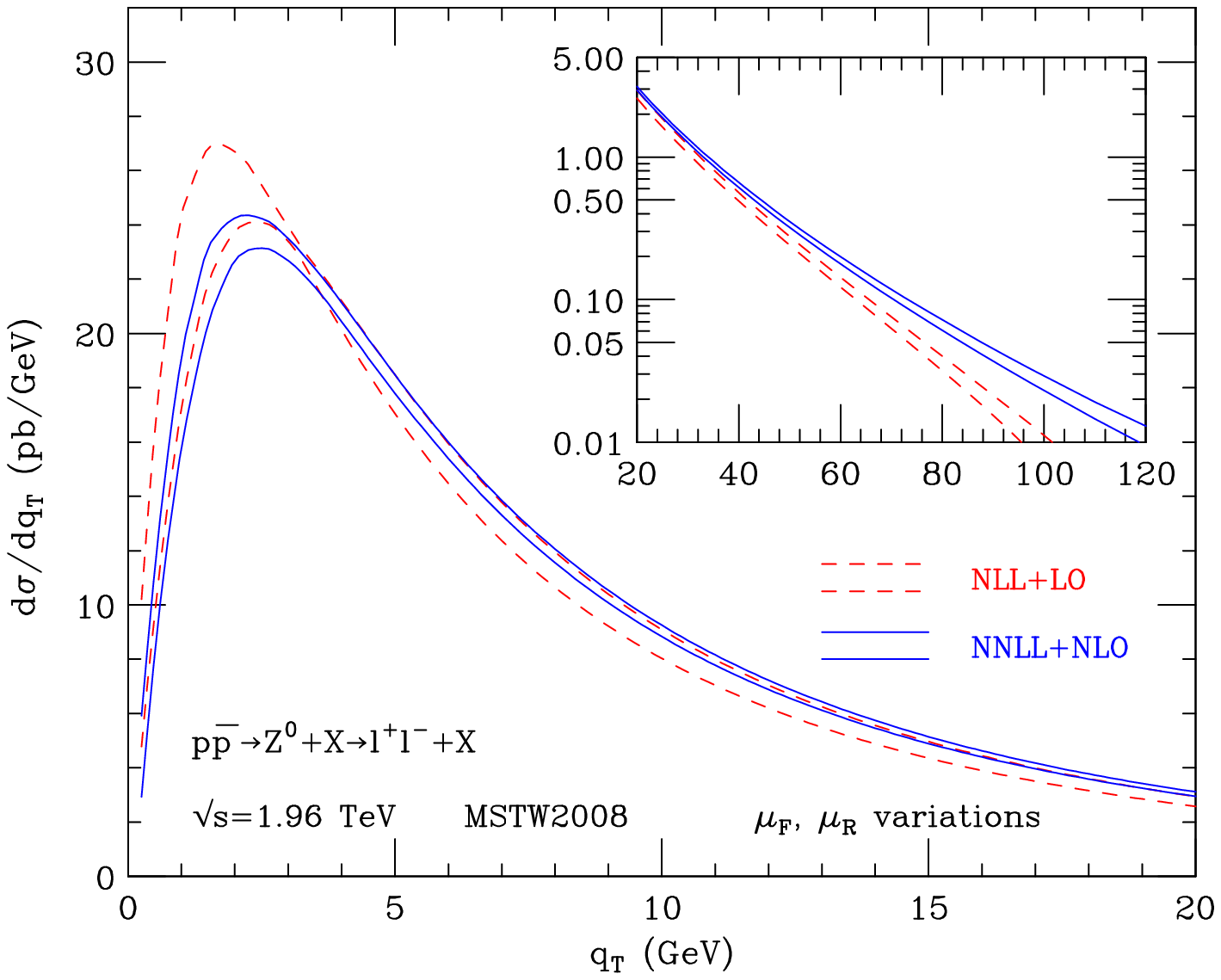} & \includegraphics[width=0.48\textwidth]{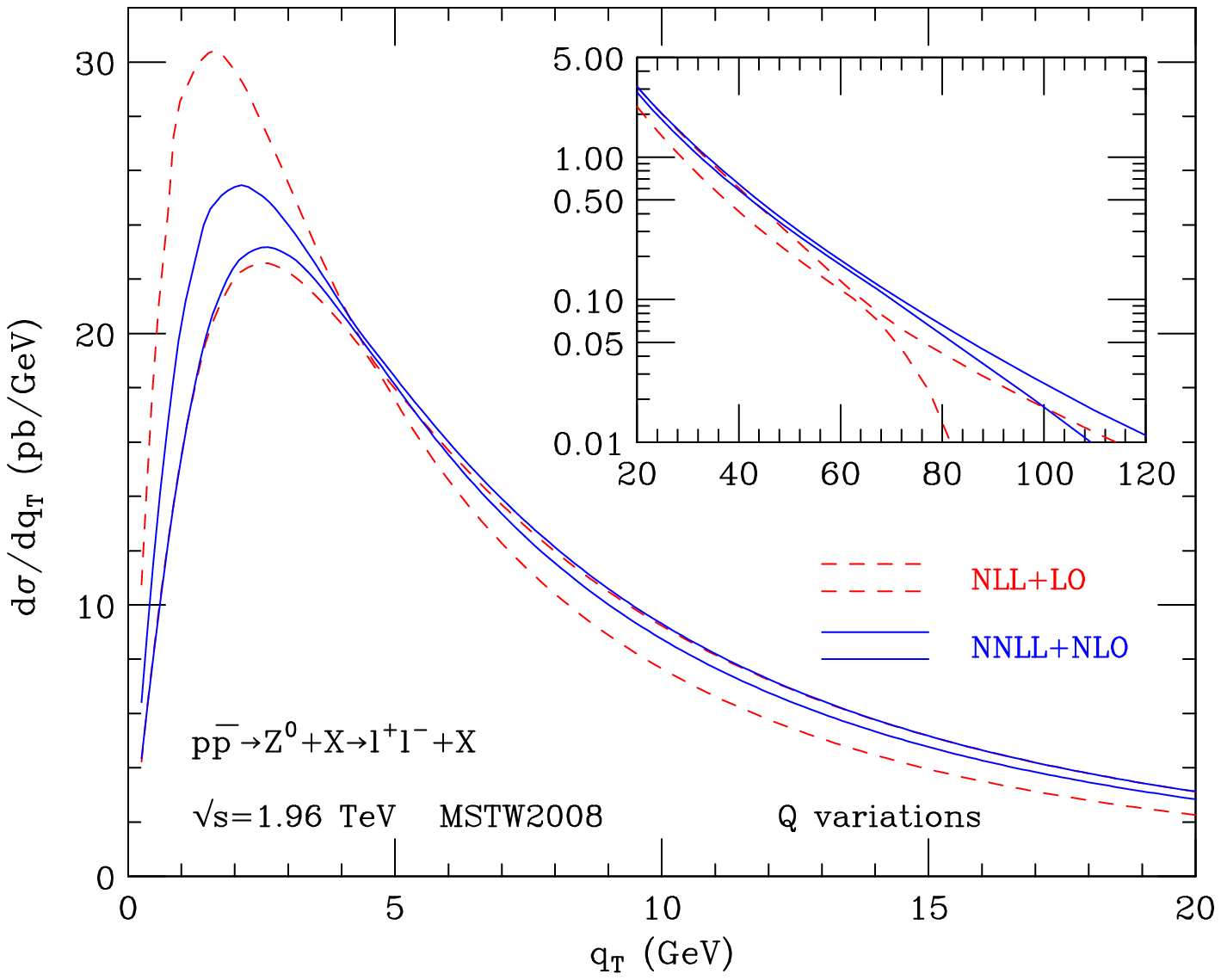}\\
\end{tabular}
\end{center}
\caption{\label{fig2}
{\em The $q_T$ spectrum of $Z$ bosons at the Tevatron Run II. The bands are obtained by varying $\mu_F$ and $\mu_R$ (left panel) and $Q$ (right panel) as described in the text.}}
\end{figure}
 
In Fig.~\ref{fig2} we 
show the scale dependence
of the NLL+LO (dashed lines) and NNLL+NLO (solid lines) results.
In the left panel we consider variations of the renormalization and factorization scales. 
The bands are obtained by varying $\mu_R$ and $\mu_F$ as 
previously described in this section.
We note that, in the region of small and intermediate transverse momenta
($q_T\ltap 30$~GeV), the NNLL+NLO and NLL+LO 
bands
overlap. This feature, which is not present in the case of the fixed-order
perturbative results at LO and NLO (see Figs.~2 and 3 in Ref.~\cite{Bozzi:2008bb}), confirms the importance
of 
resummation to achieve a stable perturbative prediction.
In the region
of small and intermediate values of $q_T$,
the main difference between the NNLL+NLO and NLL+LO predictions
is in the size of the scale 
variation
bands. 
Going from NLL+LO to NNLL+NLO accuracy,
we observe a reduction
of the scale dependence from  $\pm 4$\% to $\pm 3$\% at the peak,
from $\pm 7$\% to $\pm 3$\% at $q_T\sim 20$~GeV, and from $\pm 7$\% to $\pm 5$\% at $q_T\sim 50$~GeV.
We point out that the $q_T$ region where resummed perturbative predictions are
definitely significant is a wide region from intermediate to
relatively-small (say, close to the peak of the distribution) values of 
$q_T$.
In fact, at very
small values of $q_T$ (e.g. $q_T\ltap 5$~GeV) the size of non-perturbative effects is
expected to be important,
while in the
high-$q_T$ region (e.g. $q_T\gtap 60$~GeV) the resummation of the logarithmic terms
cannot improve the predictivity of the fixed-order perturbative expansion.
The inset plot in the left panel of Fig.~\ref{fig2} shows the region
from intermediate to large values of $q_T$. At large $q_T$, the NLL+LO and NNLL+NLO
results deviate from each other, and the deviation increases as $q_T$ increases.
As previously 
stated,
this behaviour is not particularly worrying since,
in the large-$q_T$ region, the resummed results
loose their predictivity and can (should) be replaced by customary  
fixed-order results.

In the right panel of Fig.~\ref{fig2} we consider resummation scale variations. The bands are obtained
by fixing $\mu_R=\mu_F=m_Z$ and varying $Q$ between $m_Z/4$ and $m_Z$.
Performing variations of the resummation scale,
we can get further insight on the size of yet uncalculated 
higher-order logarithmic contributions at small and intermediate values of 
$q_T$.
We find that the 
scale dependence
at NNLL+NLO (NLL+LO)
is about $\pm 5$\% ($\pm 12$\%) in the region of the peak,
and about $\pm 5$\% ($\pm 16$\%) in the region where $q_T\sim 20$~GeV.
We note that in 
a wide region of $q_T$ values,
$5$~GeV$\ltap q_T\ltap 50$~GeV, the resummation scale dependence is  
reduced by, roughly, a factor of 2 in going from the NLL+LO to the NNLL+NLO result.
Comparing the left and right panels of Fig.~\ref{fig2},
we see that, at NNLL+NLO accuracy,
the resummation scale dependence is larger than (though, comparable to)
the $\mu_F$ and $\mu_R$ dependence.

The integral over $q_T$ of the resummed NNLL+NLO (NLL+LO)
spectrum is in agreement (for any values 
of $\mu_R, \mu_F$ and $Q$) with the value of
the corresponding NNLO (NLO) total cross section to better than 1\%,
thus checking
the numerical accuracy of our code. 
We also note that the large-$q_T$ region gives a little contribution to
the total cross section (see some numerical results in Sect.~3.2 of Ref.~\cite{Bozzi:2008bb});
therefore, the total cross section
constraint mainly acts as a perturbative constraint on the resummed 
spectrum 
in the region from intermediate
to small values of $q_T$.

\begin{figure}[htb]
\begin{center}
\begin{tabular}{cc}
\includegraphics[width=0.48\textwidth]{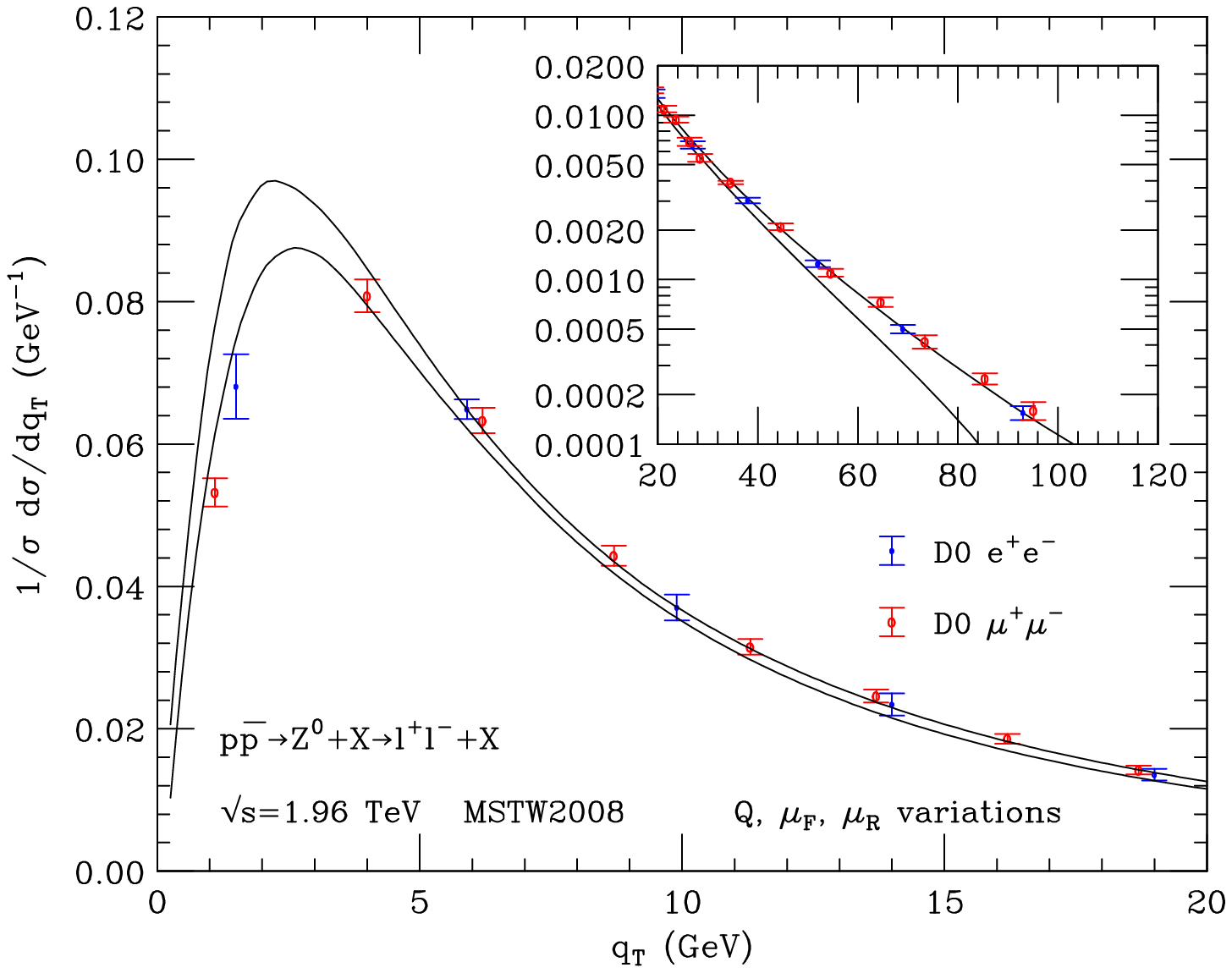} & \includegraphics[width=0.48\textwidth]{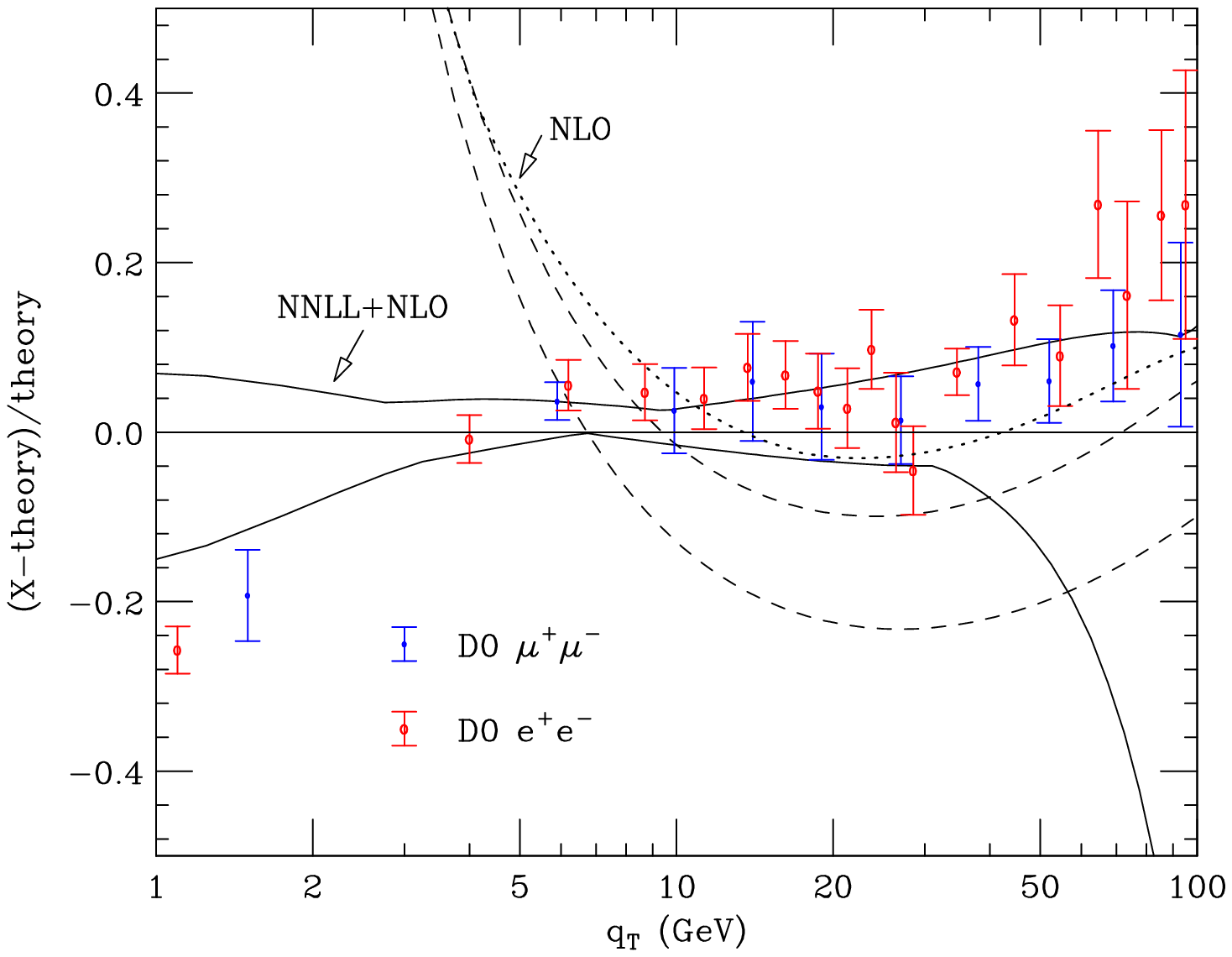}\\
\end{tabular}
\end{center}
\caption{\label{fig3}
{\em The normalized $q_T$ spectrum of $Z$ bosons at the Tevatron Run II. The
NNLL+NLO result is compared with the D0 data of Refs.~\cite{:2007nt,Abazov:2010kn}. 
The bands are obtained as described in the text.}}
\end{figure}

The D0 Collaboration has 
measured 
the normalized $q_T$ distribution,
$\frac{1}{\sigma}\frac{d\sigma}{dq_T}$, from data at the Tevatron Run II in the 
$e^+e^-$ 
\cite{:2007nt}
and 
$\mu^+\mu^-$
\cite{Abazov:2010kn} channels.
In the left panel of Fig.~\ref{fig3} 
we report the D0 data and our corresponding results at NNLL+NLO accuracy.
The NNLL+NLO band 
represents
our estimate of the perturbative uncertainty, and it is obtained by performing
scale variations as follows. We independently vary $\mu_F,\mu_R$ and $Q$
in the ranges
$m_Z/2\leq \{\mu_F,\mu_R\} \leq 2m_Z$ and $m_Z/4\leq Q\leq m_Z$,
with the constraints $0.5 \leq \mu_F/\mu_R \leq 2$ and $0.5 \leq Q/\mu_R \leq 2$.
The constraint on the ratio $\mu_F/\mu_R$ is the same
as used in the left panel of Fig.~\ref{fig2}; it has the purpose of avoiding
large logarithmic contributions (powers of $\ln(\mu^2_F/\mu^2_R)$)
that arise from the evolution of the parton densities.
Analogously, the constraint on the ratio $Q/\mu_R$
avoids large logarithmic contributions (powers of $\ln(Q^2/\mu_R^2)$)
in the perturbative expansion of the resummed form factor\footnote{We do not apply additional constraints
on the ratio $Q/\mu_F$,
since the form factor
does not depend on $\mu_F$.} $\exp\{{\cal G}_N\}$
(see Eq.~(\ref{exponent})).
We recall (see e.g. Eq.~(19) of Ref.~\cite{Bozzi:2005wk})
that
the exponent ${\cal G}_N$ of the form factor is obtained by $q^2$ integration
of perturbative functions of $\as(q^2)$ over the range $b_0^2/b^2 \leq q^2\leq Q^2$.
To perform the integration with systematic logarithmic accuracy,
the running coupling $\as(q^2)$ is then expressed in terms of $\as(\mu_R)$ (and $\ln(q^2/\mu_R^2)$).
As a consequence, the renormalization scale $\mu_R$ should not be too different from the
resummation scale $Q$, which controls the upper bound of the $q^2$ integration.

The D0 data and the NNLL+NLO band are presented in the left panel of
Fig.~\ref{fig3}.
The inset plot shows the region from $q_T=30$~GeV up to $q_T=100$~GeV.
A quick inspection of the figure
shows that the data are described quite well by 
the NNLL+NLO perturbative predictions.

Differences and similarities between theoretical calculations and the data
are more clearly visible by considering their fractional difference with respect
to a 'reference' theoretical result.
We choose the NNLL+NLO result at central values of the scales
(i.e. $\mu_F=\mu_R=m_Z$, $Q=m_Z/2$) as `reference' theory, and we 
show
the ratio
(X$-$theory)/theory in
the right panel of Fig.~\ref{fig3}. The label $X$ refers to either the experimental data
or the NNLL+NLO (solid lines) and NLO (dashed lines and dotted line) results, 
including their scale dependence. 

Considering the right panel of Fig.~\ref{fig3},
we first 
comment on the scale uncertainty band
of the NNLL+NLO result (solid lines).
Such uncertainty is about $\pm$6\% at the peak, it decreases to about
$\pm$4--5\% in the region up to $q_T=10$~GeV, and then it 
increases, reaching the size of 
about $\pm$12\% at $q_T=50$~GeV. 
In the region beyond $q_T\sim 60$~GeV the resummed
result looses predictivity, and its perturbative uncertainty 
becomes large.
The right panel of Fig.~\ref{fig3} also shows the 
scale variation
band of the NLO result.
The NLO band (dashed lines) is 
obtained by varying $\mu_F$ and $\mu_R$ 
(the NLO calculation does not depend on the resummation scale $Q$)
as in the NNLL+NLO calculation.
We comment on the comparison between the NLO and NNLL+NLO bands.
At large values\footnote{The available D0 data in the region 
100~GeV$< q_T < 250$~GeV are consistent with the NLO result 
(see Refs.~\cite{:2007nt, Abazov:2010kn} 
and Fig.~5 in Ref.~\cite{Bozzi:2008bb}).} of $q_T$,
the NLO and NNLL+NLO bands overlap (the NLO and NNLL+NLO are certainly consistent),
and the NLO result has a smaller uncertainty.
At intermediate values of transverse momenta,
the NLO result is lower than the NNLL+NLO result,
and the corresponding scale variation bands do not overlap. 
We recall (see the discussion in
Sect.~3.1 of Ref.~\cite{Bozzi:2008bb}) that
in this region the NLO band underestimates 
the true perturbative uncertainty of the NLO result;
indeed, the NLO band and the corresponding LO band do not overlap at
intermediate values of $q_T$ (see Figs.~3 and 5 in Ref.~\cite{Bozzi:2008bb}).
To get some quantitative insight into the `true' perturbative uncertainty
of the NLO calculation in this $q_T$ region, we can consider wider scale
variations and, in particular, we can lower the values of
$\mu_F$ and $\mu_R$.
In the right panel of Fig.~\ref{fig3}, we show the NLO band that we 
obtain by independently varying $\mu_F$ and $\mu_R$ in the range
$m_Z/4\leq \{\mu_F,\mu_R\} \leq 2m_Z$,
with the constraint $0.5 \leq \mu_F/\mu_R \leq 2$:
this band is delimited by the dotted line and the 
lower dashed line
(the region between the dotted line and the central values of the dashed
band roughly corresponds to scale variations in the range
$m_Z/4\leq \{\mu_F,\mu_R\} \leq m_Z$). 
We note that lowering the scales at NLO improves the 
consistency between the NLO and NNLL+NLO results.
We also note that we have considered similar enlarged scale variations at 
NNLL+NLO accuracy, and we have checked that they do not significantly modify
the NNLL+NLO band in the intermediate-$q_T$ region. This confirms the better
stability of the NNLL+NLO calculation with respect to scale variations.
In the small-$q_T$ region, the 
NLO result is theoretically unreliable.
The NLO band quickly deviates from the NNLL+NLO band as $q_T$ decreases.

The right panel of Fig.~\ref{fig3} shows that
the NNLL+NLO result is consistent with the D0 data, even at very low values of
$q_T$. We note that the resummed result is obtained in a 
perturbative
framework. At low values of $q_T$, non-perturbative effects are important
and are expected (see, e.g., 
the final part of Sect.~5 in Ref.~\cite{Bozzi:2008bb}) to 
shift
the resummed result such as to improve the agreement with the data.
In the region where $q_T\ltap 50$~GeV, the experimental errors and the corresponding NNLL+NLO errors
overlap, with the sole exception of a couple of data points at very low $q_T$.
In the same region, the perturbative uncertainty of the NNLL+NLO result turns out to be
comparable with the size of the experimental errors.
As pointed out by the D0 Collaboration \cite{:2007nt,Abazov:2010kn},
the NLO result tends to undershoot the data in the region of 
intermediate values of $q_T$: NNLL resummation improves 
the agreement with the data in this $q_T$ region. 

\section{Summary}
\label{sec:summa}

In this paper we have considered the $q_T$ spectrum 
of DY lepton pairs produced in hadron collisions, and we have presented a 
perturbative QCD study
based on transverse-momentum resummation at the NNLL order.

We have followed the formalism developed in 
Refs.~\cite{Catani:2000vq,Bozzi:2005wk}, which is valid for the production of a
generic high-mass system of non strongly-interacting particles 
in hadron collisions. 
The formalism combines small-$q_T$ resummation at a given logarithmic accuracy 
with the fixed-order calculations. It implements a unitarity constraint
that guarantees that the integral over $q_T$ of the differential cross section
coincides with
the total cross section at the corresponding fixed-order accuracy.
This leads to QCD
predictions with a controllable and uniform perturbative 
accuracy
over the region from small up to large values of $q_T$. 
At large values of $q_T$, the resummation formalism is superseded by customary
fixed-order calculations.

We have considered the explicit case of 
DY lepton
pairs from the decay of
a $Z$ boson produced at the Tevatron Run~II.
Using the recently computed NNLL
coefficient ${\cal H}_N^{V(2)}$~\cite{Catani:2009sm},
we have extended the NLL+LO resummed calculation presented in 
Ref.~\cite{Bozzi:2008bb} to the NNLL+NLO accuracy.
The NNLL corrections are not large and make the $q_T$ spectrum slightly harder.
We have performed 
a study of the scale dependence of the calculation
to estimate 
the corresponding perturbative uncertainty. 
In a wide region of transverse momenta ($5$~GeV$\ltap q_T\ltap 50$~GeV) 
the size of the scale uncertainties is
considerably reduced in going from
NLL+LO to NNLL+NLO accuracy.

We have compared the resummed calculation with the results of measurements 
\cite{:2007nt, Abazov:2010kn} of  
the normalized $q_T$ spectrum 
at the Tevatron Run~II. 
The perturbative uncertainty of the NNLL+NLO results turns out to be
comparable with the experimental errors.
The NNLL+NLO 
results (without the inclusion of any non-perturbative effects)
are consistent with the experimental data in a wide region of
transverse momenta.
Comparing the NNLL+NLO and NLO results,
we have also shown that NNLL resummation 
improves the
agreement with the data at intermediate values of $q_T$.
As is well known (and theoretically expected), the NLO result fails to describe 
the data at small values of $q_T$.

More detailed comparisons with available data on vector boson production
and further studies of theoretical uncertainties, including
the impact of non-perturbative effects, are left to future investigations.

\vspace*{3mm}
\noindent {\em Note added}. After the completion of this paper,
the value of the coefficient $A^{(3)}$ (see the related comment in the
final part of Sect.~\ref{sec:theory}) for $q_T$ resummation was derived in 
Ref.~\cite{Becher:2010tm}. We have checked the quantitative effect of 
this value of $A^{(3)}$ on our results for the $q_T$ distribution of 
$Z$ bosons. We find that the effect is generally very small.
The largest effect is produced in the region of very low values of $q_T\,$;
for instance, in the case of $Z$ production at the Tevatron, the quantitative
effect is at the level of about 2\% (4\%) at $q_T \simeq 2$~GeV (1~GeV).

\end{document}